\documentclass{PoS}
\usepackage{graphicx}
\usepackage{amsmath,amsfonts}

\newcommand{\Tr}{\textrm{Tr}}
\newcommand{\ehat}{\hat{e}}
\newcommand{\ev}[1]{\left\langle #1 \right\rangle}

\title{Isotriplet Dark Matter on the Lattice: \\ SO(4)-gauge theory with two Vector Wilson fermions}

\ShortTitle{SO(4) gauge theory with two fermions}

\author{\speaker{Ari Hietanen}\\
        E-mail: \email{hietanen@cp3-origins.net}}

\author{Claudio Pica\\%\footnote{CP$^3$-Origins \& the Danish Institute for Advanced Study DIAS,
        E-mail: \email{pica@cp3-origins.net}}

\author{Francesco Sannino$^\dagger$\\
        E-mail: \email{sannino@cp3-origins.net}}

\author{Ulrik S{\o}ndergaard$^\dagger$\\
        E-mail: \email{sondergaard@cp3-origins.net}}

\author{\\CP$^3$-Origins \& the Danish Institute for Advanced Study DIAS,
        University of Southern Denmark, Campusvej 55, DK-5230 Odense M, Denmark.}

\abstract{We present preliminary results for simulations of SO(4)-gauge theory with two Dirac Wilson fermions transforming according to the vector representation. We map out the phase diagram including the strong coupling bulk phase transition line as well as the zero PCAC-mass line. In addition, we measure the pseudo scalar and vector meson masses, and investigate whether the theory features chiral symmetry breaking. If the theory is used for breaking the electroweak symmetry dynamically it is the orthogonal group equivalent of the Minimal Walking Technicolor model but with the following distinctive features: a] It provides a natural complex weak isotriplet of Goldstone bosons of which the neutral component can be identified with a light composite dark matter state; b] It is expected to break the global symmetry spontaneously; c] It is free from fermionic composite states made by a techniglue and a technifermion.\\\\
CP3-Origins-2012-027 and DIAS-2012-28}

\FullConference{The 30 International Symposium on Lattice Field Theory - Lattice 2012,\\
		June 24-29, 2012\\
		Cairns, Australia}

\begin{document}

\section{Introduction}

{ Understanding the phase diagram of strongly interacting theories will unleash a large number of theories of fundamental interactions useful to describe the very fabric of the bright and dark side of the universe \cite{Sannino:2008ha}.  To gain a coherent understanding of strong dynamics  besides the SU(N) gauge groups \cite{Sannino:2004qp,Dietrich:2006cm}, one should also investigate the orthogonal, symplectic and exceptional groups.
SO(N) and SP(2N) phase diagrams were investigated with analytic methods in \cite{Sannino:2009aw}, while the exceptional ones together with orthogonal gauge groups featuring spinorial matter representations were studied in \cite{Mojaza:2012zd}. So far lattice simulations have been mostly employed to explore the phase diagram of SU(N) gauge theories while a systematic lattice analysis of the smallest symplectic group  was launched in \cite{Lewis:2011zb}. Here we move forward by analyzing on the lattice the dynamics of the SO(4) gauge group with two Dirac fermions in the vector representation of the group. 
The choice of this specific orthogonal gauge theory is based on the following theoretical and phenomenological considerations: 
\begin{itemize}
\item It is expected to be below or near the lower boundary of the conformal window \cite{Sannino:2009aw,Frandsen:2009mi}. 
\item If used for technicolor \cite{Weinberg:1979bn,Susskind:1978ms} the simplest choice of the hypercharge assignment compatible with gauge and Witten anomalies leads to integer electric charges for the composite states. In Minimal Walking Technicolor \cite{Sannino:2004qp,Dietrich:2005jn,Foadi:2007ue}, for example, one achieves integer charged composite states but also to doubly charged heavy leptons. 
\item The technicolor theory leads to a weak isotriplet with the neutral member an ideal dark matter candidate \cite{Sannino:2009aw,Frandsen:2009mi}, the  (Isotriplet Technicolor Interactive Massive Particle) iTIMP. This state is a pseudo Goldstone and therefore can be light with respect of the electroweak scale making it an ideal candidate to resolve some of the phenomenological and experimental puzzles \cite{Frandsen:2009mi}.  The first model featuring composite dark matter pions appeared in \cite{Gudnason:2006ug,Ryttov:2008xe} and the first study of technipion dark matter on a lattice appeared in \cite{Lewis:2011zb}.

\end{itemize}}

Due to the reality of the fermion representation the quantum global symmetry group  is SU(4) expected to break spontaneously to SO(4), yelding nine Goldstone bosons. Once gauged under the electroweak theory three are eaten by the SM gauge bosons. Six additional Goldstone bosons form an electroweak complex triplet of technibaryon with the neutral isospin zero component to be identified with the iTIMP of \cite{Frandsen:2009mi}.

SO(4) is a semi simple group, SO(4) $\cong$ SU(2)$\otimes$SO(3), and it has a non-trivial center $Z_2$. The two-loop $\beta$-function of the theory does not have an infrared fixed point. The theory is asymptotically free and can display chiral symmetry breaking. However, we would like to confirm this with lattice simulations, since there is also the possibility that the theory is walking \cite{Sannino:2009aw,Frandsen:2009mi,Holdom:1981rm,Yamawaki:1985zg,Appelquist:1986an}.

Previous studies of the pure gauge part behavior for orthogonal groups on the {\it lattice} were done in \cite{deForcrand:2002vs}. As a natural first step, we have studied the phase diagram in the ($\beta,m_0$)-plane to find the relevant region of parameter space to simulate. We have determined the zero PCAC-mass line as well as the strong coupling bulk phase transition line. In addition, we have studied pseudo scalar and vector meson masses and found indication of chiral symmetry breaking.
\section{Lattice study}
We use Wilson plaquette action with Wilson fermions on the lattice:
\begin{align}
  S   & = S_g + S_f \\
  S_g & = \beta \sum_{x,\mu\ne\nu}\left(1-\Tr\left[ U_\mu(x)U_\nu(x+\ehat_\mu)U^\dagger_\mu(x+\ehat_\nu)U^\dagger_\nu(x)\right]\right) \\
  S_f & =  - \frac12 \sum_{x,y,\mu}\bar\psi(x)\left[\left(1-\gamma_\mu\right) U_\mu(x)\delta_{x,y-\ehat_\mu}  + \left(1+\gamma_\mu\right) U^\dagger_\mu(x-\ehat_\mu)\delta_{x,y+\ehat_\mu} \right]\psi(y),
\end{align}
where $\beta=8/g^2$ is the lattice coupling, $m_0$ is the bare fermion mass and $U_\mu$ are the real link variables.

\begin{figure}
  \begin{center}
    \includegraphics[width=0.49\textwidth]{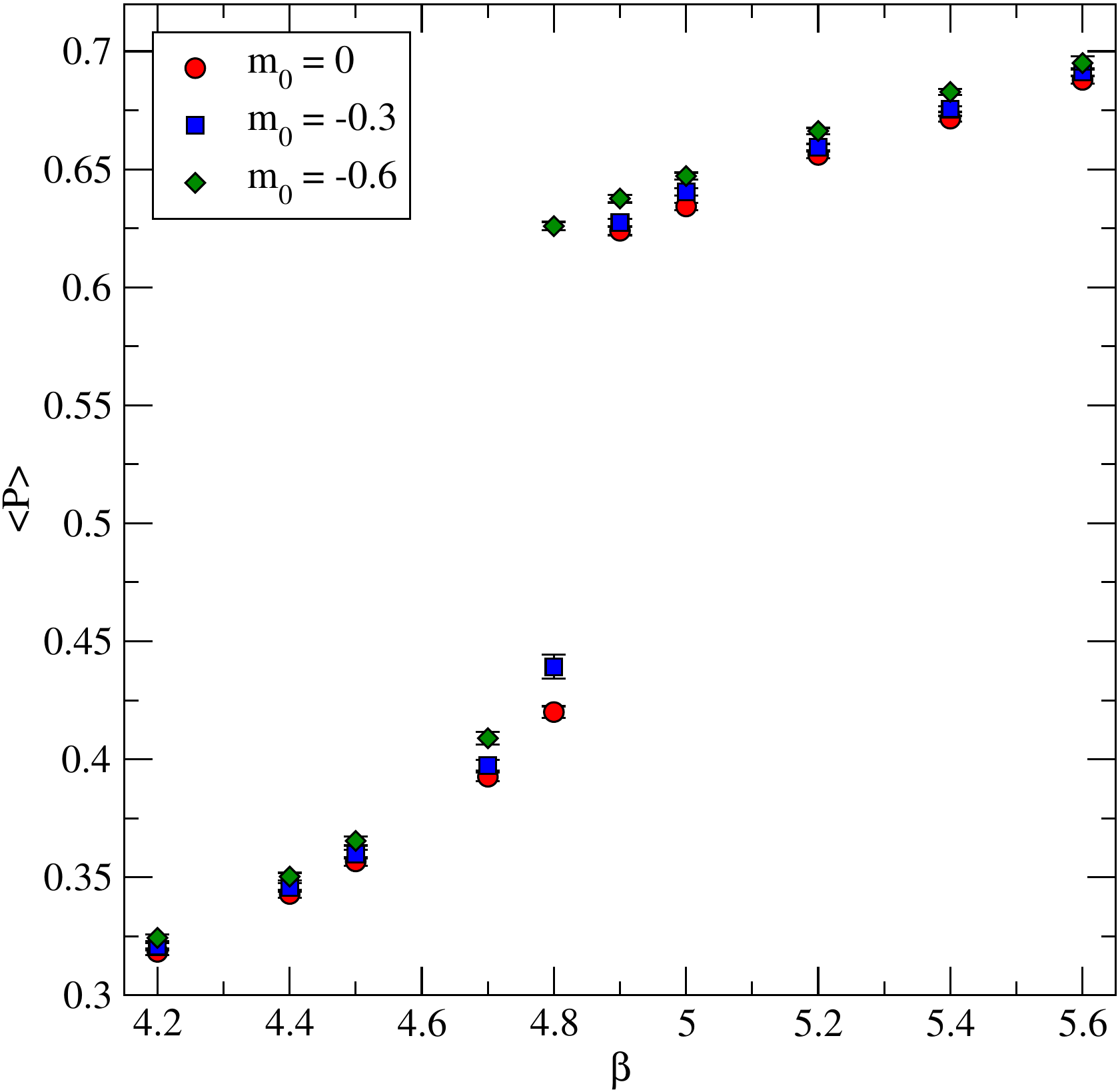}
    \includegraphics[width=0.49\textwidth]{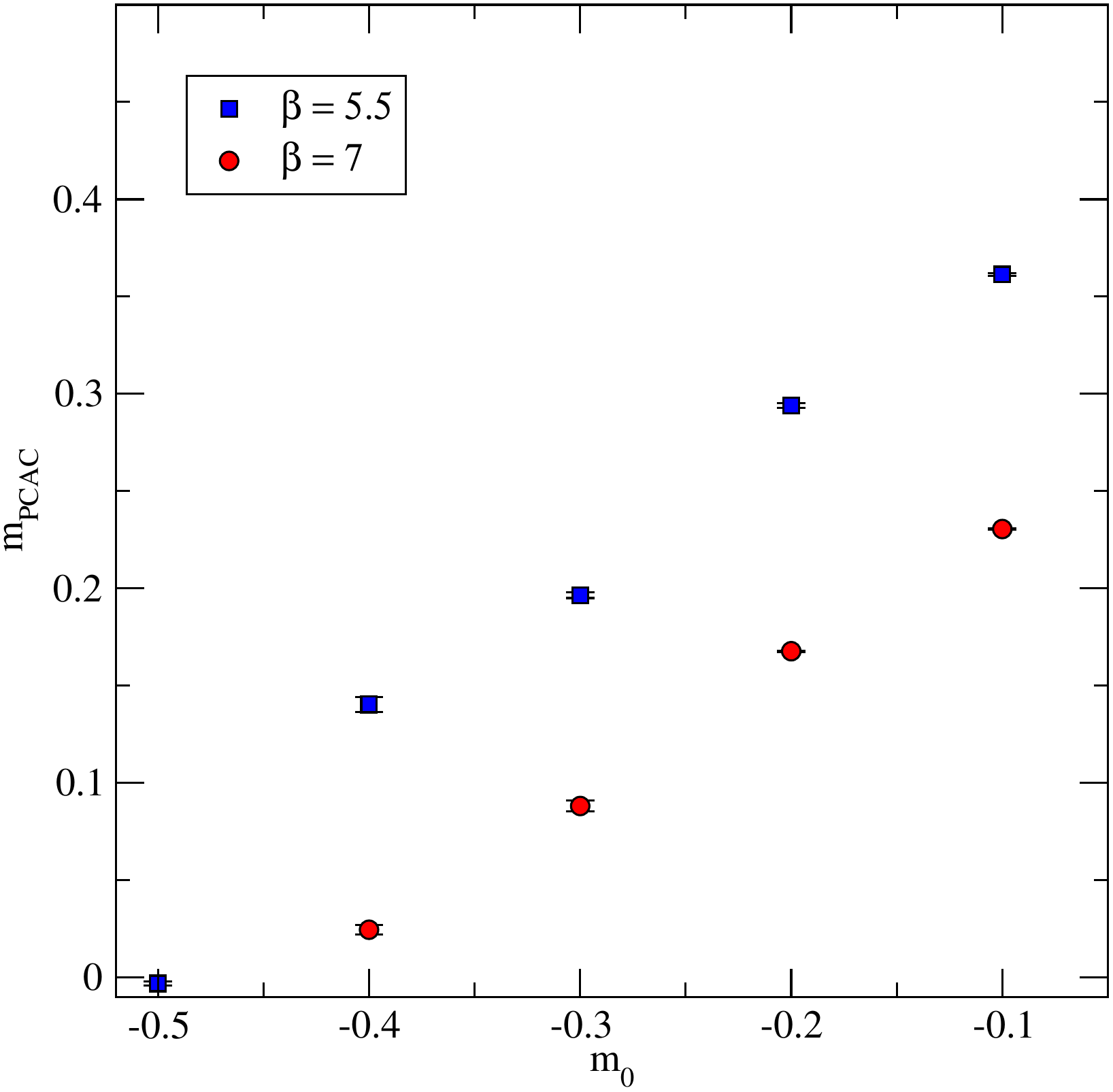}
    \caption{Left: The plaquette expectation value as a function of the lattice coupling $\beta$ for different bare masses. The jump around $\beta=4.8$ indicates the strong coupling bulk phase transition. Right: PCAC-mass quark mass as a function of bare mass. The lattice volume is $V=16\times8^3$ in both panels.\label{bulkphase}}
  \end{center}
\end{figure}

\begin{figure}
  \begin{center}
    \includegraphics[width=0.7\textwidth]{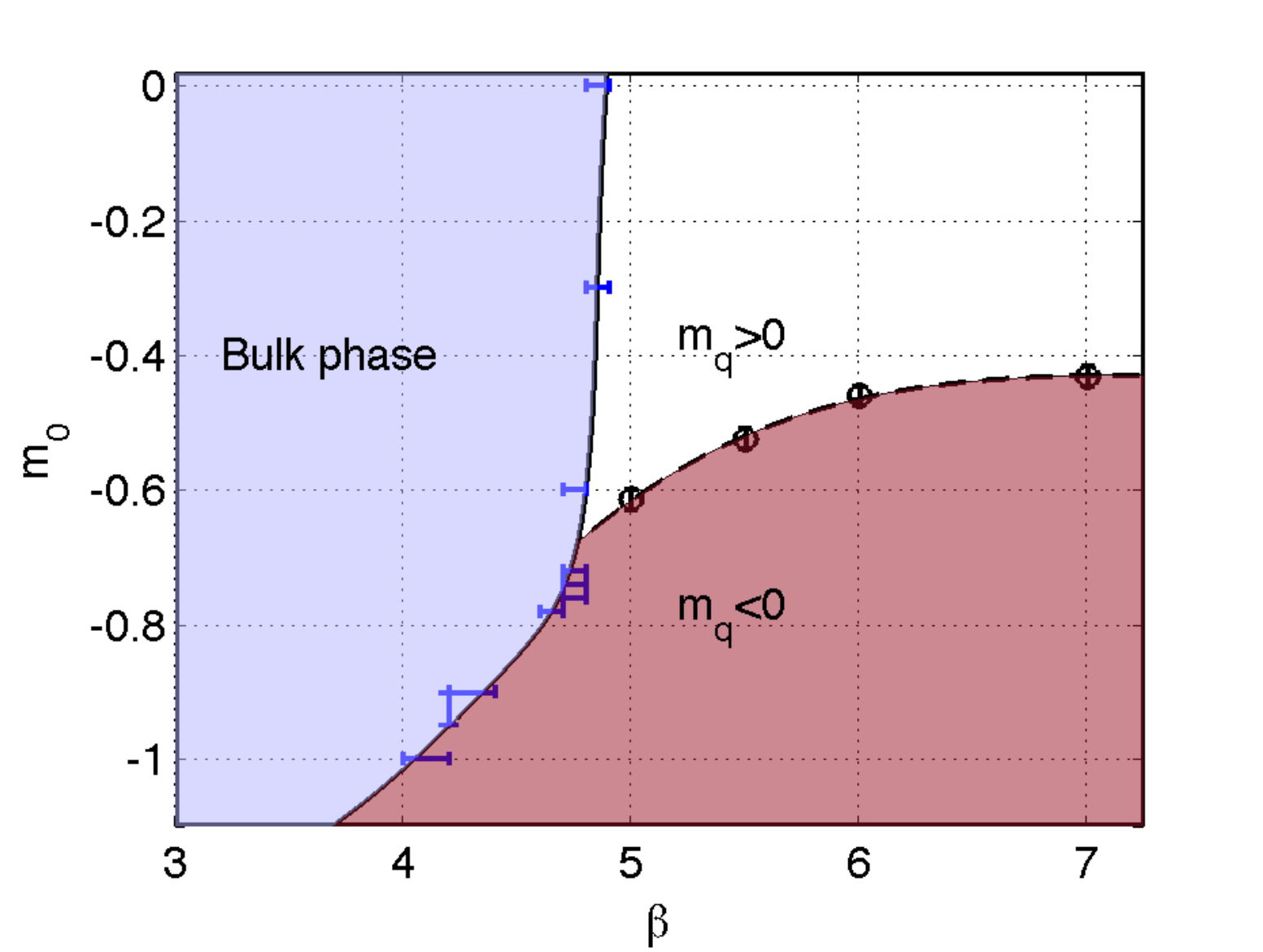}
    \caption{The phase diagram showing the bulk phase transition and the zero fermion mass line for $V=16\times8^3$.\label{phases}}
  \end{center}
\end{figure}

To map out the interesting region of parameter space, we have performed simulations on relatively small volumes $V=16\times8^3$. The bulk phase transition can be easily located by a discontinuity in the plaquette expectation value. See left panel of Fig.~\ref{bulkphase}. 

The fermion mass is measured using the axial Ward identity (PCAC mass):
\begin{equation}
  m_{\rm PCAC}=\lim_{t \rightarrow \infty}\frac{1}{2}\frac{\partial_t V_{\rm PS}}{V_{\rm PP}},
\end{equation}
where the currents are
\begin{align}
  V_{\rm PS}(x_0) &= a^3\sum_{x_1,x_2,x_3} \ev{\bar{u}(x)\gamma_0 \gamma_5 d(x)\bar{u}(0)\gamma_5d(0)}\nonumber \\
  V_{\rm PP}(x_0) &= a^3\sum_{x_1,x_2,x_3} \ev{\bar{u}(x)\gamma_5d(x)\bar{u}(0)\gamma_5d(0)}.
\end{align}
We are especially interested in finding the critical line, along which the fermion mass vanishes. For a small mass extrapolation see right panel of Fig~\ref{bulkphase}. The resulting phase diagram is given in Fig.~\ref{phases}

\begin{figure}
  \begin{center}
    \includegraphics[width=0.49\textwidth]{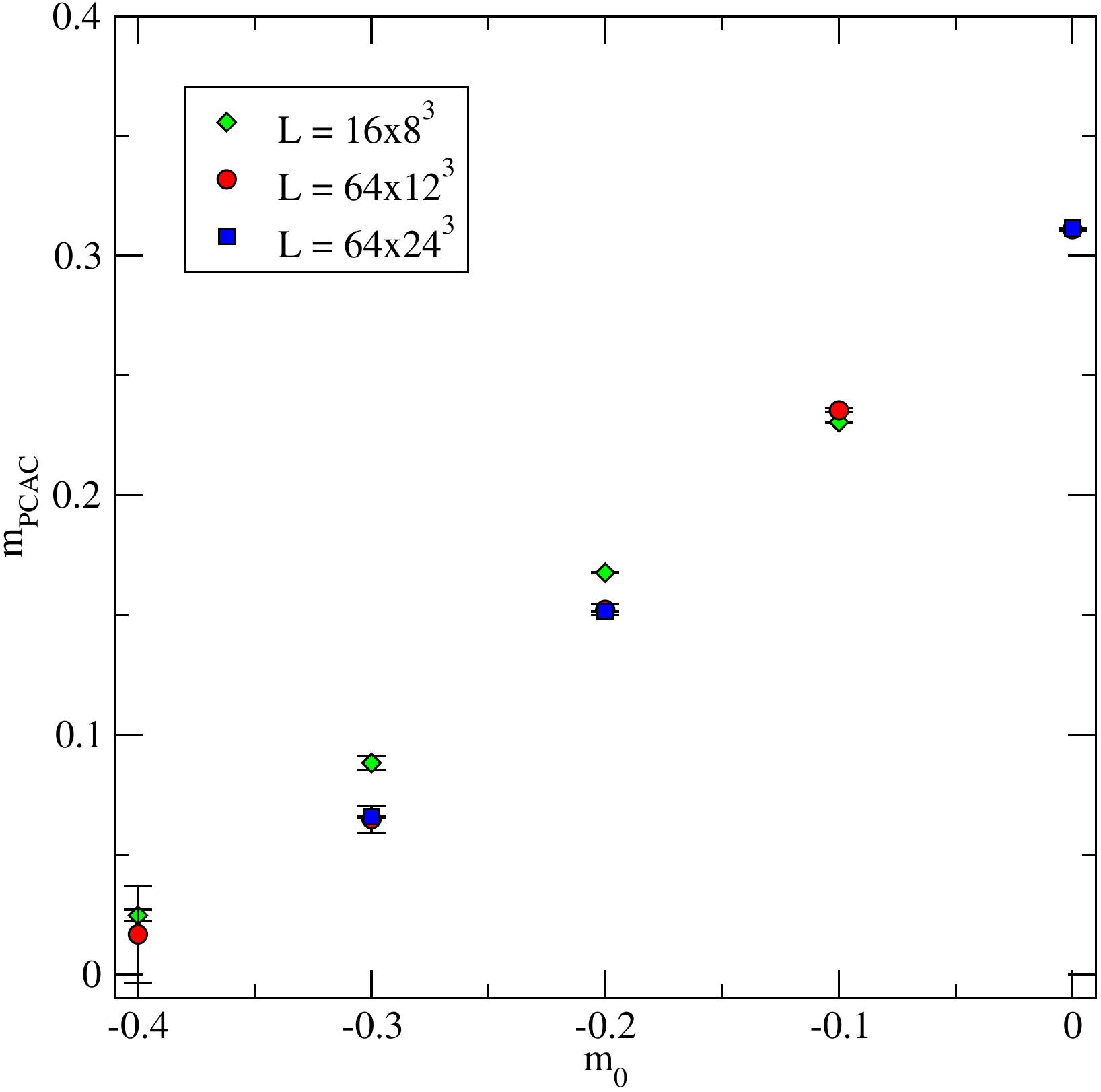}
    \includegraphics[width=0.49\textwidth]{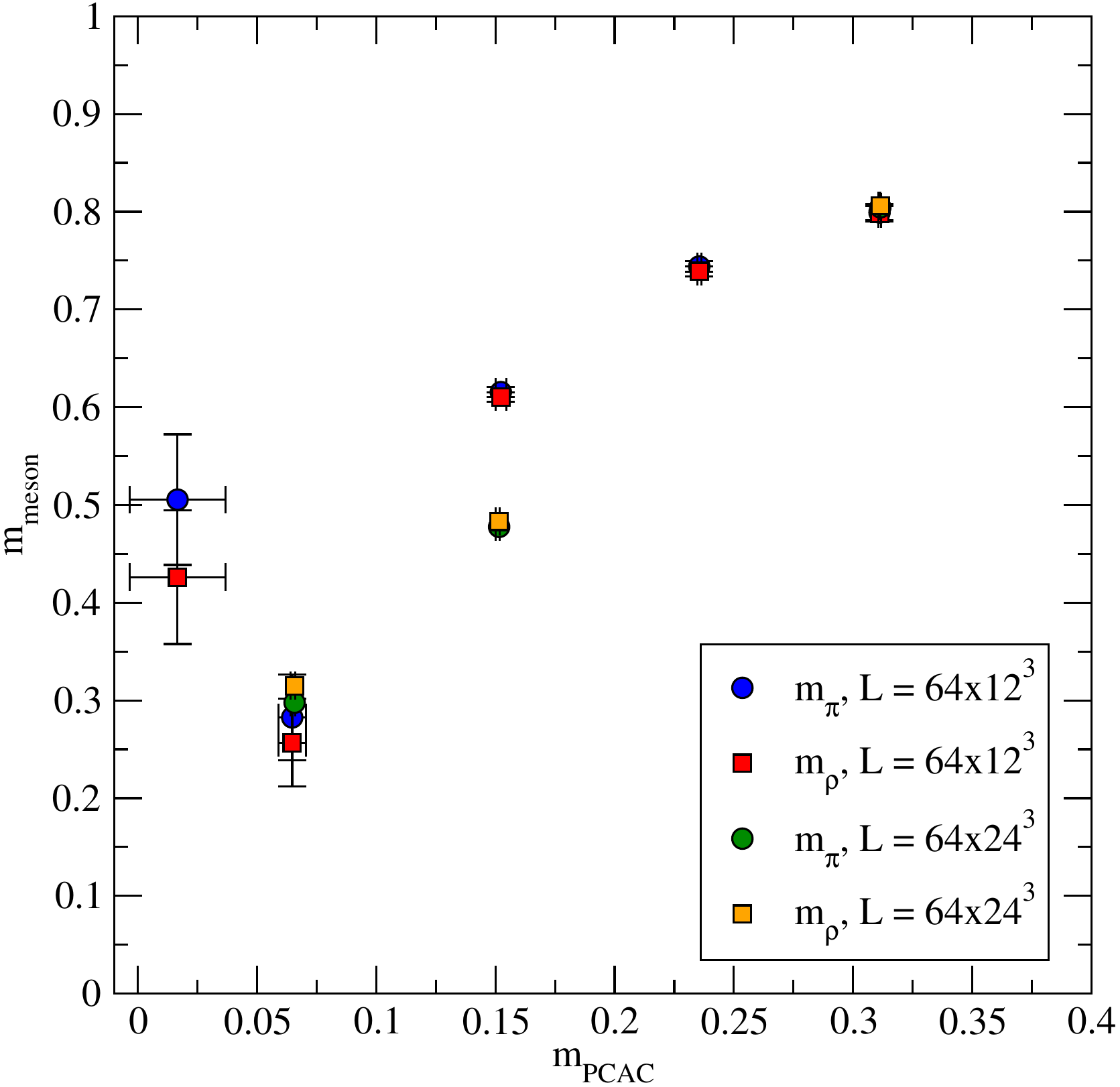}
    \caption{Left: PCAC-mass as a function of bare mass. The finite volume effects are under control. Right: Pseudo scalar and vector meson mass as a function of PCAC-mass. At low masses the finite volume effects are large, and smaller volume, $V=64\times12^3$, results are plagued with lattice artefacts. \label{m0mpcac}}
  \end{center}
\end{figure}

The meson masses are estimated calculating time slice averaged zero momentum correlators 
\begin{equation}
C_{\bar u d}^{(\Gamma )}(t)= \sum_{\bf x,y}\Tr \left( \left[  \bar u(t,{\bf x}) \Gamma d(t,{\bf x}) \right]^\dagger  \bar u(0,{\bf y}) \Gamma d(0,y)  \right) \, ,
\end{equation}
where $\Gamma=\gamma_5$ for pseudo scalar meson and $\Gamma=\gamma_k$ ($k=1,2,3$) for vector meson. 

To study the meson spectrum and the possible chiral symmetry breaking phenomenon we performed simulations also on larger volumes $V=64\times12^3$ and $V=64\times24^3$ with $\beta=7$. The chosen value of $\beta$ is far away from the bulk phase transition, but still at a relatively large coupling. In the left panel of Fig.~\ref{m0mpcac} we have plotted the PCAC-mass as a function of bare fermion mass. The smallest volume $V=16\times 8^3$ seems to suffer slightly from finite volume effects, but the simulations with two larger volumes lay on top of each other.

However, the meson masses do not fare so well. In the right panel of Fig.~\ref{m0mpcac} the estimates of pseudo scalar and vector meson masses are plotted for the two largest volume. At the highest PCAC-mass the values agree, but already at relatively high PCAC-mass $\sim 0.15$ there is a significant deviation between simulations of $V=64\times12^3$ and $V=64\times24^3$. As one approaches the chiral limit, the simulations with $V=64\times12^3$-lattice are not consistent. For example, the meson masses increase as the PCAC-mass is lowered. We think that this behavior is related to a breaking of a center-symmetry in regions of space\footnote{The fermion term explicitly breaks the center symmetry, but one can still distinguish phases where the Polyakov-loop is large or small.}. 

\begin{figure}
  \begin{center}
    \includegraphics[width=0.49\textwidth]{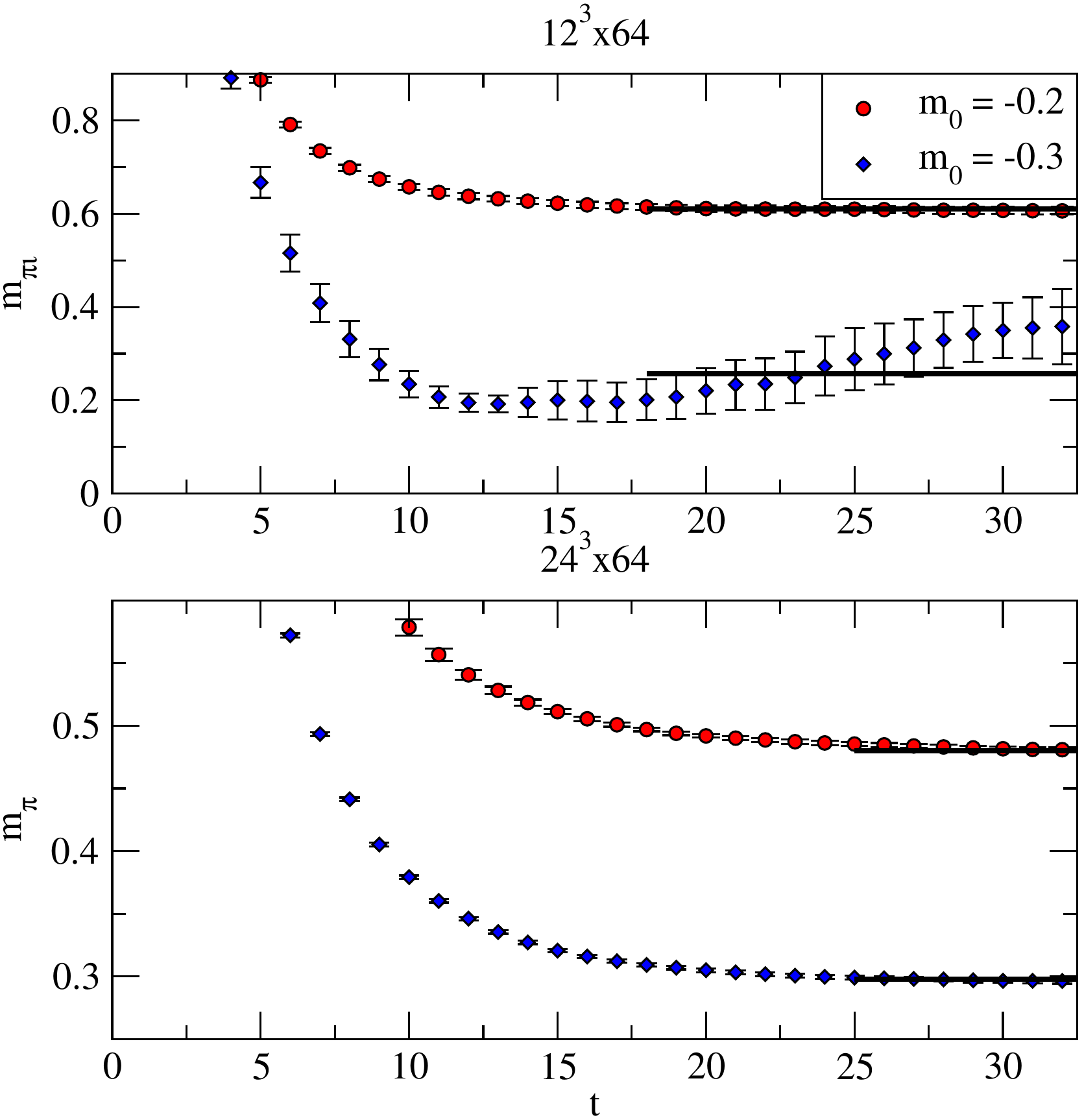}
    \includegraphics[width=0.49\textwidth]{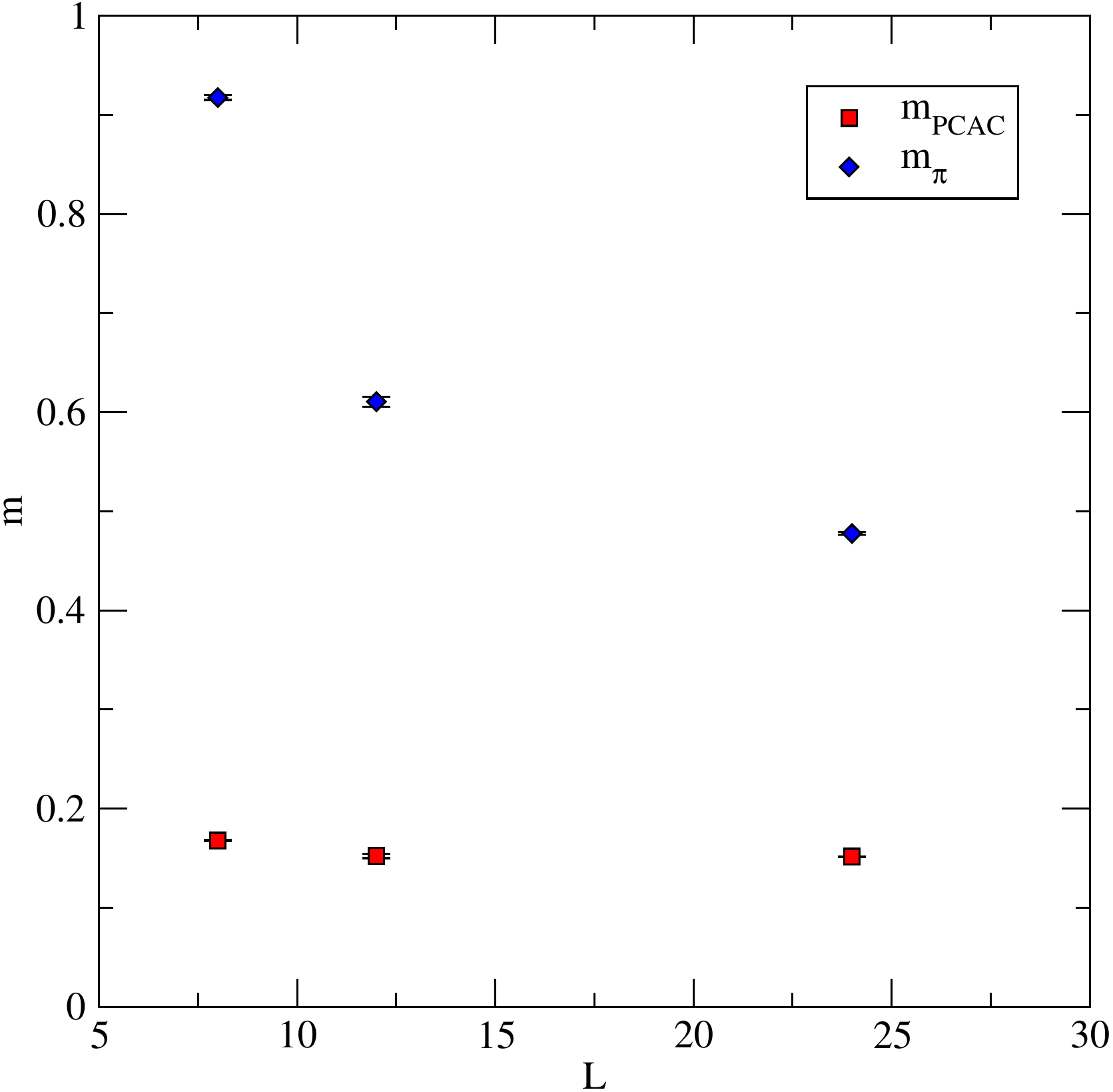}
    \caption{Left: The effective mass plot for with two different volumes for $\beta=7$. Right: PCAC-mass and a pseudo scalar meson mass as a function of spatial lattice size for bare mass $m_0=0.2$ and $\beta=7$. \label{finitevol}}
  \end{center}
\end{figure}

The phenomenon is studied in more detail in \cite{Hietanen:2012xx}, but the effective mass plots demonstrates the problem. In the left panel of Fig.~\ref{finitevol}, we have plotted the pseudo scalar effective mass for two values of bare mass ($m_0=-0.2$ and $-0.3$) and volume ($V=64\times12^3$ and $V=64\times24^3$) as a function of the temporal distance $t$. The masses of the mesons are obtained from the plateau at the large $t$. For the small volume with lighter fermions there is no plateau to fit the mass, but a unexpected rise in the effective mass. This is due to metastable phases which might arise in the lattices with small volumes. Therefore, we do not think that the smaller volume simulations are related to the continuum values. However, the phenomena is easily observed from the correlator, which behave well for larger volume, and we believe that $V=64\times24^3$ lattice is large enough to produce continuum physics results.

%% The phenomenon is studied in more detail in \cite{Hietanen:2012xx}, but the effective mass plots demonstrates the problem. In the left panel of Fig.~\ref{finitevol}, we have plotted the pseudo scalar effective mass for two values of bare mass ($m_0=-0.2$ and $-0.3$) and volume ($V=64\times12^3$ and $V=64\times24^3$) as a function of the temporal distance $t$. The masses of the mesons are obtained from the plateau at the large $t$. For the smaller volume with lighter fermions there is no plateau to fit the mass, but a unexpected rise in the effective mass. This is due to two different phases, which are characterized by the different Polyakov-loop expectation value, in the system. The specturm of the theory is different in the different phases which results in an unusual behavior of the effective mass. Therefore, we do not think that the smaller volume simulations with light masses are related to the continuum values. However, the phenomena is easily observed from the correlator, which behave well for larger volume. Hence, we believe that $V=64\times24^3$ lattice is large enough to produce continuum physics results.

In the right panel of Fig.~\ref{finitevol} we have plotted the PCAC-mass and pseudo scalar meson mass as the function of spacial extent of the lattice for $m_0=-0.2$. As noted earlier the finite volume effects are mild in PCAC-mass, but quite large on the meson masses. 
\begin{figure}
  \begin{center}
    \includegraphics[width=0.7\textwidth]{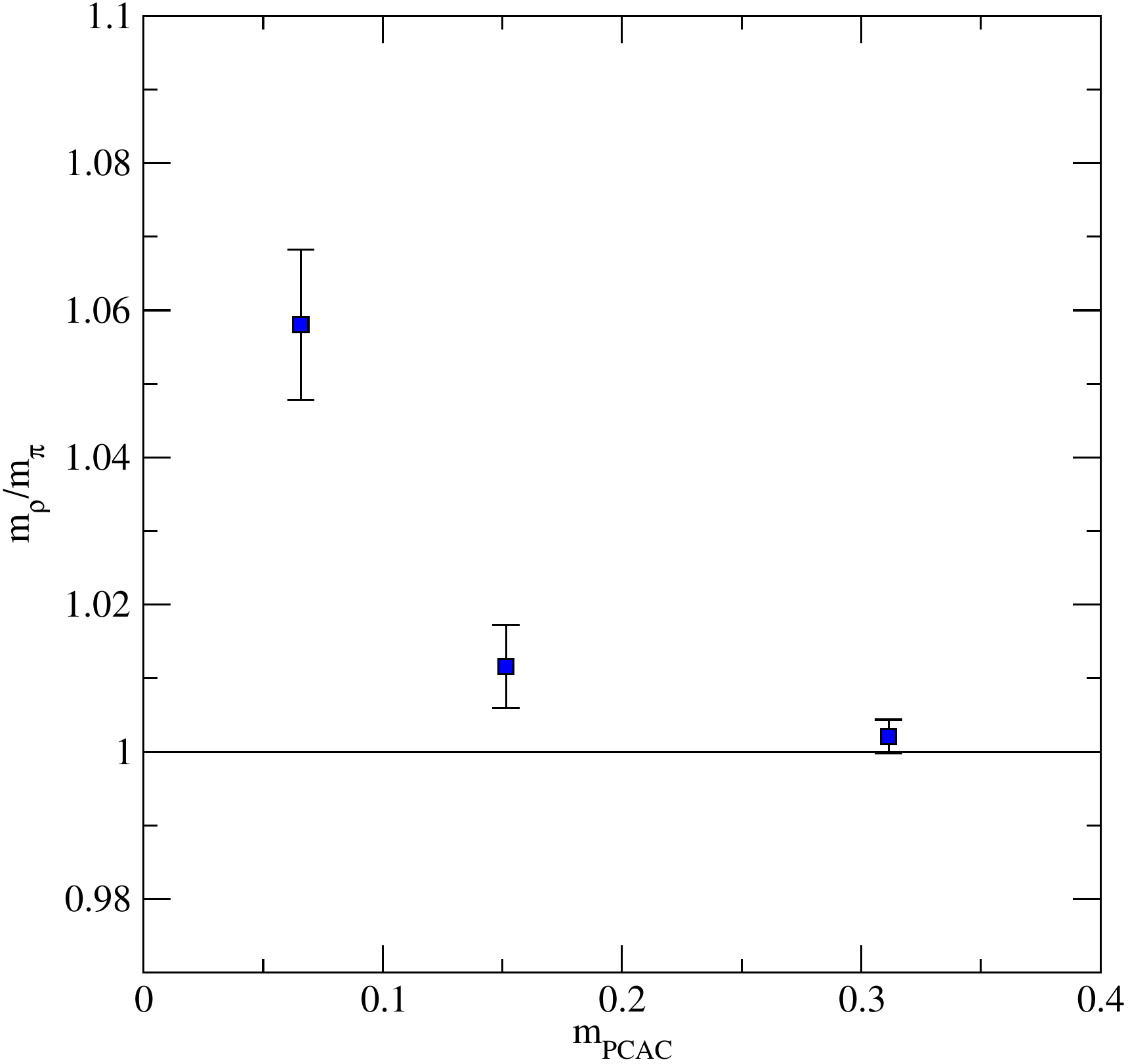}
    \caption{The mass of the vector meson divided by pseudo scalar meson. The increase at the chiral limit is an indication of chiral symmetry breaking.\label{csbreak}}
  \end{center}
\end{figure}

If the theory is conformal the meson masses depend linearly on the quark masses, whereas in the case of chiral symmetry breaking, the Goldstone bosons become masless in the chiral limit, while the other hadrons remain massive. In addition, the Goldstone bosons should approach chiral limit as $m_\pi \sim m_{\textrm{ PCAC}}^2$. Our data are not good enough to fit the power of the pseudo scalar meson mass as the chiral limit is approached, but we can compare the ratio of pseudo scalar meson mass to vector meson mass which is expected to diverge in the chiral limit (see Fig.~\ref{csbreak}). We observe a clear increase in the ratio (vector meson/pion mass) when approaching the zero PCAC-mass indicating that chiral symmetry is breaking spontaneously. However, a more detailed analysis is needed to disentangle whether the ratio continues to increase or if it plateaus to a constant value as the PCAC mass is further decreased.

\section{Conclusions}
We have performed exploratory studies of a SO(4)-gauge theory with two Dirac  fermions transforming according to the vectorial representation of the gauge group. As in pure gauge theory there is a bulk phase transition below $\beta=5$, which depends mildly on the fermion masses. The computation of meson masses suffers from the finite volume effects and require relative large lattices. We believe that lattices of size $V=64\times24^3$ are needed to produce meaningful results. We find preliminary indications of chiral symmetry breaking, but a conclusive result requires  further studies. 
\acknowledgments
The numerical calculations presented in this work have been performed on the Horseshoe5 and Horseshoe6 clusters at the University of Southern Denmark (SDU) funded by the Danish Center for Scientific Computing for the project ``Origin of Mass'' 2009/2010.

\end{document}